# BioJam Camp: toward justice through bioengineering and biodesign co-learning with youth

**Authors:**

Callie Chappell[1], Henry A.-A.[†], Elvia B. O.[†], Emily B.[†], Bailey B.[†], Jacqueline C.-M.[†], Caroline Daws[1], Cristian F.[†], Emiliano G.[†], Pagé Goddard[2], Xavier G.[†], Anne Hu[3,†], Gabriela J.[†], Kelley Langhans[1], Briana Martin-Villa[4], Penny M.-S.[†], Jennifer M.[†], Soyang N.[†], Melissa Ortiz[5,6], Aryana P.[†], Trisha S.[†], Corinne Takara[6,7], Emily T.[†], Paloma Vazquez[8], Rolando Perez[6,*], Jen Marrero Hope[9,*]

**Affiliations:**

[1] Department of Biology, Stanford University, Stanford, CA 94305, USA

[2] Department of Genetics, Stanford University, Stanford, CA 94305, USA

[3] School of Engineering, Tufts University, Medford, MA 02155, USA

[4] Department of Bioengineering, Stanford University, Stanford, CA 94305, USA

[5] Industrial Design, California College of the Arts, San Francisco CA 94107, USA

[6] Xinampa, Salinas, CA 93905, USA

[7] Okada Design/Nest Makerspace, Cupertino, CA 95014, USA

[8] Program in Human Biology, Stanford University, Stanford, CA 94305, USA

[9] Undergraduate Research, Office of the Vice Provost for Undergraduate Education, Stanford University, Stanford, CA 94305, USA

[†] BioJam youth collaborator

*These individuals share in senior authorship

**Author Note:** Authors who are high school students at the time of writing are listed by first name and last initial(s).

**Abstract:**

BioJam is a political, artistic, and educational project in which Bay Area artists, scientists, and educators collaborate with youth and communities of color to address historical exclusion of their communities in STEM fields and reframe what science can be. As an intergenerational collective, we co-learn on topics of culture (social and biological), community (cultural and ecological), and creativity. We reject the notion that increasing the number of scientists of color requires inculcation in the ways of the dominant culture. Instead, we center cultural practices, traditional ways of knowing, storytelling, art, experiential learning, and community engagement to break down the framing that positions these practices as distinct from science. The goal of this work is to realize a future in which the practice of science is relatable, accessible, and liberatory.

In our piece, we: describe the pedagogy, values, and practices that have shaped and reshaped BioJam since the project's start in 2019; share the experiences of our teen collaborators in BioJam by showcasing their original works, including art, prose, and poetry; and share how programs like BioJam can emerge in a variety of cultural and geographic contexts, inviting readers to join our political, artistic, and scientific praxis.

# Main text

## Sobre Las Cuadrillas

BY JAQUELINE C.-M. (BioJam teen and teen mentor)

Casi no se escuchan de ellos,
Pero deben saber para ser sinceros.

La gente campesina se agachan,
La gente campesina cortan las hojas de brócoli,
La gente campesina siembran y crean frutas y vegetales,
La gente campesina lo empacan todo,
La gente campesina alimenta América.

Las reglas del campo recompensan más por las horas que trabajen.

Y después entra la pandemia,
Que nomás hace su trabajo más laborioso.

Con solo una mascarilla reusable para todo el año.
Vienen al trabajo enfermos porque la falta de trabajo les hace mas daño.

Les pagan menos si se enferman y dejan de trabajar,
Pero tienen que luchar.

Luchar por sus sueños,
Luchar por sus familias,
Luchar por la renta,
¡Luchar en alimentar a toda la gente!

Son luchadores.

DEDICADA PARA LA CUADRILLA 254

## About the Crews

They are hardly heard of,
But you must know them to be honest.

Field workers bend down,
Field workers cut broccoli leaves,
Field workers plant and grow fruits and vegetables,
Field workers pack and sort everything,
Field workers feed America.

The rules of the field pay more for more hours they work.

And then the pandemic comes,
Making their job much more laborious.

Provided with only one reusable mask for the whole year.
They come to work sick since missing work hurts them even more.

They get paid less if they get sick and stop working,
But they need to fight.

Fight for their dreams,
Fight for their families,
Fight for their rent,
Fight to feed all of America!

They are fighters.

DEDICATED TO CREW 254

This poem was first published in Grow by Ginkgo

Social and environmental justice cannot be disentangled. Solutions to economic inequality, adaptation to a changing climate, and dismantling systems of oppression must start locally, through community (*1*). Through stories, we share knowledge: from the agricultural fields of the Salinas Valley, to protesting religious persecution in Tibet, biology, culture, and art ties together our survival.

Biodesign and bioengineering, where the natural world serves as inspiration for innovation, is heralded as a technocratic solution to emerging challenges such as improving agricultural practices, addressing healthcare inequity, and reducing greenhouse gas emissions (*2*). And yet, cultural, local, lived and Indigenous knowledge have always been central to designing around and with biology (*3*).

Bioengineering and biodesign must be by the people, for the people. However, the voices who control the narrative of science--particularly in bioengineering and biodesign--do not represent our society's breadth of identities and cultural diversity (*4*, *5*). To contribute to a more sustainable and just world, we must amplify the voices, ideas, and solutions of youth from communities that have been excluded from the conversation. Youth carry wisdom learned from their elders into the future: in this way, youth are conduits of knowledge for their communities.

BioJam is a political, artistic, and educational project in which scientists and artists collaborate with youth and communities of color to re-envision the future of bioengineering and biodesign. Our biodesign praxis centers cultural practices, traditional ways of knowing, storytelling, art, experiential learning, and community engagement to realize a future in which bioengineering is relatable, accessible, and liberatory.

Centered in the Greater Bay Area of California, BioJam's radical mission is not to teach, but to co-learn with youth through their own creativity and culture. Now in our third year, the program starts with a summer camp where instructors co-learn with teens on the topics of biodesign, biohacking, speculative design, and restorative justice (*6*, *7*). After camp ends, the program continues throughout the academic year, as youth design and lead projects to engage their home communities in what they have learned. BioJam works to develop deep roots across learning spaces: universities, museums, community bio labs, community gardens, and after school programs. These new intersectional and intergenerational learning spaces push us to expand what is understood as science in the future of bioengineering and biodesign.

*Our mission*

Our mission is to create a space that centers existing science knowledge from communities and cultures excluded from mainstream scientific practice. In the Western colonialist tradition of scientific practice, legitimacy has predominantly been offered by exclusionary institutions of higher education. However, the system of higher education, especially in the United States, privileges the voices of those for whom universities and colleges were built -- largely white, male, affluent communities. This history of science forces us to ask critically: Who creates knowledge and who legitimizes or canonizes it? Who benefits from the mainstream conception of science? Who is excluded? What knowledge, technological advances, and intergenerational best practices are lost or delayed as a result of this exclusion? BioJam centers the existing knowledge held in our cultures and centers our communities as producers and holders of knowledge. We hope that this mission opens a conversation with dominant-culture scientific practitioners in a range of educational spaces about the necessity of not just including or consulting, but elevating community-held knowledge in the process of scientific and artistic exploration and innovation. To achieve our most innovative solutions to contemporary challenges, we must acknowledge the harmful history of oppression of community-derived knowledge in scientific practice and education. Community-derived knowledge takes many forms outside of the dominant cultural practice of knowledge creation and legitimation, such as Traditional Ecological Knowledge, story-telling, experiential knowledge from working the land (*8*), imaginative futures through fiction (*9*), and creative problem solving with community-sourced materials outside of a lab space (*10*, *11*). Innovation in science and art rooted in justice must embrace community-derived knowledge in ways that deeply engage issues of gender, race, class, age, ability, and oppression.

*Our praxis*

In light of the historical exclusivity of knowledge creation and legitimation in elite social spheres and educational institutions, it is critical that the work that we do in BioJam be nested within our teens' communities. As specific scientific practices, biodesign and biotechnology require us to look both forward and backward and acknowledge the trajectories of sustainability and exploitation. Communities historically excluded from scientific practice in formal institutions have long created and passed down traditional knowledge of sustainability and creative use of biomaterials. Rather than generating participatory knowledge with teens by immersing them in institutions created by the dominant culture, such as universities or museums, BioJam seeks to empower teens to create knowledge within their home communities. Expanding the conception of biotechnology, BioJam seeks to build community-embedded spaces for biotechnological innovation and biodesign, especially to address challenges teens have identified in their communities. Beginning with youth, we uplift and center the perspectives of teens on the future of bioengineering, biodesign, and sustainability. To do this, we center pedagogies of co-design, co-learning, radical reuse (*12*, *13*), and liberatory design (*14*). By inviting teens to engage with STEAM (science, technology, engineering, arts, and math) through their own place of expertise, rooted *in* their communities, we aim to empower them to reorient and expand science by co-designing STEAM programming *for* their communities. When youth engage their own communities in ways that are culturally meaningful, they also expand intergenerational engagement and contribute to the long histories of knowledge sharing and legitimation within their family and community culture. When dominant culture practitioners engage with BioJam, they broaden their own understanding of what science can be and have the opportunity to co-design creative ways of solving important problems outside of the dominant scientific paradigm.

*Our positionality*

BioJam is more than an educational program or a camp, we are an intergenerational community of learners. We are youth, and we are elders. We are scientists, and we are artists. We are educators, and we are students. We are colonizers, and we are colonized. We are activists. We embrace our contradictions, our divergences, and use them to guide us to solutions inconceivable by dominant cultural practice.

The authors of this paper represent the breadth of participation in BioJam: high school student participants, student educators, community advisory board members, Stanford students and staff, and the founders of BioJam. Together we steward a vision for a just, radical future of bioengineering and biodesign that centers lived experience, cultural knowledge, and art, critically interwoven with dominant-culture methodologies. The conception of this paper and the construction of the outline, drafts, figures, and editing have been done collectively during meetings, using a virtual whiteboard and Google Docs, and have centered intergenerational mentoring.

Many of us are immigrants or children of immigrants. Many come from agricultural communities: from Hawaii, the American Midwest, South, and Pacific Northwest, and the Salinas Valley of California. To BioJam, we bring stories of queerness, of religious practice, of carcerality, and precarity. We also bring histories of imperialism, neoliberalism, and cultural genocide. The process of growing an organization that seeks true collaboration between organizers and youth from low-resource communities and Stanford University feels impossible at times. Perhaps it is. As such, the confluence of our positionalities sometimes creates friction, sometimes transcendence.

BioJam was founded in 2019 during a fateful rideshare with Corinne Okada Takara, a community bioartist and educator, and Dr. Rolando Perez, a PhD bioengineer. Rolando and Corinne discovered that they had an overlapping interest in developing STEM/STEAM programming that engages communities through their own cultural histories and expertise. They began to incubate ideas for a camp that would introduce teens to bioengineering and creative biomaterial explorations while also celebrating cultural arts. In the three years since, BioJam has connected youth from more communities around Northern California, enmeshed into the Xinampa community bio lab in Salinas, CA, created a student organization at Stanford, and developed deep community ties through advisory boards.

Through Corinne and Rolando's vision, we have centered our work in the Greater Bay Area of Northern California. We seek to center lived experience into cultural, historical, and sociopolitical narratives. Through this, we often uncover tensions centered in place. BioJam participants and educators work multiple jobs to survive, an effect of income inequality propagated by Stanford University, the explosion of the tech industry in Silicon Valley, and agriculture and other industries in the Greater Bay Area. Where we are able to live is shaped by a history of redlining and racial violence, and yet BioJam participants and organizers span the most and least wealthy zip codes in the Bay Area. We and our families are sick, exposed to chemicals used by conventional agriculture, oil refineries and industrial waste, disproportionately affecting communities of color; yet, we also work in scientific laboratories that produce such contamination. Our land burns, our lungs fill with smoke, and our rivers fill with soot from wildfires that are exacerbated by neoliberal policies. And yet, we are all consumers.

We seek to find solidarity in tension. We strive to create a community where mutual learning, aid, and vulnerability can give rise to interconnectedness, regeneration, and flourishing (*15*). Through solidarity, we hope to find innovative solutions to protect our homes and communities. We redress the shortcomings of historical aspirations of science and craft an expansive vision for the future that balances the needs of all and provides abundant space for curiosity, creativity and play.

*Our structure*

BioJam is a complex and interwoven ecosystem of youth, activists, artists, educators, students, academics, and professionals **(Figure 1)**. From this diverse community, we come together into four different groups that drive the energy behind BioJam: youth, advisory boards, community educators, and a Stanford student organization.

**BioJam aims to be an organization by and for youth and communities, and at the heart of our decision-making processes lie the wants and needs of youth** (Figure 1a). Youth participate in BioJam in two ways. First, they join as "campers", who attend the BioJam summer camp and continue collaborating with BioJam throughout the rest of the year-long program. As campers, they co-learn and co-create science and art with the rest of the BioJam team, and their interests direct the focus of post-camp programming and community engagement projects. After their first year, youth have the opportunity to return to the program as "teen mentors." In this role, they advise on the following year's curriculum and themes and serve as mentors to the next generation of campers: leading activities, advising campers, and sharing their wisdom and experiences.

**BioJam's mission is guided by our community, teen, and academic advisory boards** (Fig. 1b)**.** We convene our advisory boards multiple times throughout the year to gain their feedback on our program, curriculum, and how we are meeting community and youth needs. Our community advisory board consists of organizations from collaborating communities, including community bio labs and innovation labs, education non-profits, a public benefits renewable energy collaborative, science museums, Indigenous land trusts, community gardens, and workforce development non-profits. These community leaders guide the focus and vision of BioJam, showing organizers how the program can meet community needs; recruiting teens to the program; connecting camp to community resources including space, funding opportunities, and internships; and providing opportunities for the teens to collaborate with them on community research projects. Many of the relationships we have built with these organizations and individuals were generously shared by Corinne Okada Takara.

Our teen advisory board is made up of former youth participants and youth mentors who provide valuable insider perspectives on the structure and themes of camp.

Our academic advisory board consists of academics from the Greater Bay Area and beyond who provide suggestions on curriculum development, assessment, and paradigmatic framing of camp. Many of the academics have ties to our teens' communities and specialize in subjects including agroecology, environmental justice, urban development, biodiversity conservation, and bioengineering.

Founding partner **[Xinampa](#)** has been crucial for BioJam to be effectively rooted in community engagement and generate successful programming. Xinampa is a community bio lab based in Salinas, CA that empowers individuals to activate, create, and innovate with biology. From the beginning, Xinampa has been the loom from which BioJam's storytelling, framing, mission, and vision has been woven. Many of BioJam's connections to community orgs are only possible through Xinampa's years of conversation, collaboration, and trust building with many of our advisory board members. BioJam's approach to programming has been influenced by Xinampa's Public Interest Technology Community Innovation Fellowship research into the needs of Salinas and South Monterey County teens (*16*). Finally, Xinampa is becoming a hub for additional STEAM opportunities for the BioJam teens during the academic year.

**BioJam would not be possible without the vision, experience, and relationships cultivated by community educators** (Fig. 1c)**.** BioJam was born from a partnership between STEAM community educator and co-author Corinne Okada Takara and bioengineer Dr. Rolando Perez. Corinne has been a guiding force behind BioJam since its inception, co-creating a vision of a STEAM co-learning experience rooted in community. First, from her decades of experience as a community-engaged artist and arts educator, Corinne dreamed up and executed innovative learning experiences that blend art and science for teen participants. Second, she has generously shared relationships with community organizations she has built over many years to create our community advisory board, as well as help us grow and cultivate those relationships. Corinne also guided us to bring on other STEAM community educators for the 2021 summer camp (see "Curriculum"), relationships we plan to deepen in future years. Corinne and these other educators are the visionaries behind our camp curriculum, guiding everything from the ethos and broad learning goals of the camp to innovativing new making and learning activities. Community educators are the threads that bring together the needs and wants of our youth, the expertise and grounding of our advisory boards, and the organizing power of Stanford collaborators. As BioJam continues to grow, we hope that our future is centered on collaboration with community educators.

**The final piece of our BioJam ecosystem is BioJam CoLABS** (Fig. 1d)**.** CoLABS (Community Learning through Art, Biology, and Solidarity) is a Stanford volunteer student organization made up of undergraduate students, graduate students, postdoctoral researchers, and staff. CoLABS members ("CoLABorators") support BioJam's organization and logistics, executing the vision created by youth and the advisory boards. BioJam CoLABS convenes advisory boards, obtains materials, secures funding and resources, facilitates curriculum creation, and organizes meetings. We take advantage of our positions in a well-resourced institution of higher education to connect other members of our ecosystem with learning opportunities, funding, materials, and public platforms that we have privileged access to. At the same time, members of CoLABS co-learn and co-create with teens, weaving our own experiences and knowledge rooted in our fields of education, biology, engineering, and computer science, as well as our own communities, into the scientific and exploratory practice that we build together with the teens and communities we work with.

*Our curriculum*

The BioJam curriculum (Fig. 3) is developed through collaboration between CoLABorators, community educators, and the community, academic, and teen advisory boards. Each year, teens envision the program's theme for the following year (September/October). Then, the curriculum is designed in iterative collaborations (November-June) that acknowledge the many roles needed for social change as described by Iyer in her reflection on social change ecosystems (*17*): visionaries and guides (community educators, advisory boards), experimenters and storytellers (teens), and weavers and builders (CoLABs). Because teens are engaged as co-designers of the curriculum, throughout their year-long journey we ask teens to think critically and imagine ways *they* would remix the curriculum for their own community.

Corinne Okada Takara has been a central visionary, developer, and guide in the BioJam curriculum. When she co-founded BioJam with Rolando Perez (then a Stanford Department of Bioengineering PhD candidate) in 2019, Corinne brought decades of experience in community-engaged art, biomaking, and education to the BioJam curriculum. Each year, she has guided our growing team to center culture, social change, and the environment.

Because the curriculum journey has changed each year, we will describe the development of the 2021 camp curriculum as an example. After teens suggested the 2021 camp focus be on environmental justice, Corinne guided us to hire three additional community educators from the community advisory board to develop curriculum modules. These educators, Leticia Hernandez, Nazshonni Brown-Almaweri, and Alba Cardenas, were experts in creating learning and co-learning spaces already in our participating teens' communities. These educators worked in a variety of institutions and programs, from youth STEAM programming (Alba Cardenas, [San Jose Alliance for Youth Achievement](#)) to community gardening (Leticia Hernandez, Local Urban Gardeners, [Center for Landbased Learning](#)) to Indigenous land rematriation (Nazshonnii Brown-Almaweri, [Sogorea Te' Land Trust)](#), and were the visionaries and guides in our curriculum development work. Importantly, they shared personal identities with our teens and were deeply embedded in each of the communities we engage with.

CoLABS team members worked to support each educators' vision and weave together co-learning experiences with teens. CoLABS built out the visions and learning goals set forward by community educators and received iterative feedback from educators, community advisory board members, and teen mentors. CoLABS members built slide decks, prototyped activities, ordered supplies and packed kits, created journal prompts, managed troubleshooting, and wove themes across camp activities. Our goal was to create curricula that centered teen knowledge and interests and create a learning environment where teens could grow their voices in communicating their communities' unique expertise and stories as central to knowledge creation and innovation.

Our learning ecosystem also included practitioners who showed us how our ideas could become reality, futurists who invited us to imagine inclusive and sustainable worlds, and community advisory board members who served as counselors in our process. For example, we invited guests to speak with campers from a wide range of cultural and professional backgrounds, including: Lois Kim, a visual strategist for NASA's Jet Propulsion Laboratory; Karina Garcia, a Salinas-born agroecology researcher; Grace-Mary Oisamoje, a Nigerian fashion designer centering social justice; and many others.

The 2021 summer camp, themed around climate change, sustainability, and environmental justice, consisted of ten camp days over two weeks. These included two days of community-building and introduction to making, three full-day modules (on Circuitry, Traditional Ecological Knowledge, and Hydroponics) with hands-on making components, four guest speaker conversations, daily journal reflections, and a final showcase with all of our community members.

As co-learners during camp, teens poured their experiences and knowledge into their work as experimenters and storytellers. Their contributions breathed life into BioJam camp, and they took activities from ideas to creative prototypes. For example, in 2021, teens were invited to decorate greenhouse/shelter structures (Fig. 2k). Their designs reflected their own life experiences – some teens attached plants of particular importance to them; others drew murals depicting the foundational members of their communities, centering agricultural workers. Teens' creative and collaborative interpretations of their own work across camp led to generative sharing of family history, cultural tales, and central values in their communities.

Empowered by their exploration of the intersections between biology, art, and engineering, BioJam teens continue collaborating throughout the academic year to plan and execute community engagement projects. Students turn their focus back to their own communities, applying newly-learned concepts to personal knowledge of their community's needs and values. While some projects are coordinated by the connections of BioJam's organizing team, others are teen-designed. BioJam's growing advisory board ecosystem provides inspiration and starting points for community projects. BioJam teen mentors support the development of these engagement projects.

*Our 2019 journey*

The BioJam program first began in 2019 as a four-day summer camp in the Uytengsu Teaching Lab in Stanford University's Department of Bioengineering (Fig. 2a). This in-person pilot program started with seven teens. In the mornings, teens learned about local and global innovation in biodesign and sustainability. Topics included learning about protein engineering, strawberry DNA extraction, operating a cell-free expression system, and bacterial cellulose synthesis. Being housed in a Stanford University lab space embedded teens in the research community and provided opportunities for them to learn about and interact with research, equipment, and academics. In the afternoons, open-ended biomaterial making activities created opportunities for the teens to bring in their creativity and perspectives. Other activities included lab work, tours and nature walks, science communication presentations, and collaborative biomaking. These explorations provided different points for the teens and mentors to share, discuss, and learn with and from each other.

The 2019 camp assessment was conducted on a large poster board with a slider scale where teens put stickers along the scale. This survey was developed in collaboration with Dr. Sonia Travaglini (Department of Mechanical Engineering, Stanford University) and was informed by cartoon surveys developed by the Grow-Your-Own (GIY) BioBuddies (three BioJam campers). On the last day of camp, we conducted two assessments: one on a large dry erase board that asked teens to add post-it notes under the titles "Liked," "Learned," "Wish," and "Surprised". Campers felt ownership over surveys conducted in the camp; they erased and rephrased the titles and added emoji images for each. The other assessment asked teens to use [Post-it Notes to respond](#) to the prompts "A favorite thing was…" "Something better could be…" and "I'd like more…" Favorite things included hands-on learning, the wide variety of materials we worked with, and meeting everyone. In the "something that could be better" category, several mentioned the need for better information and help with the pre-camp online training. In the category of "I'd like more," several mentioned the desire for 3D printing training, presentations from people pursuing PhDs, icebreakers before camp, and learning more about new innovations in bioengineering.

Feedback from both teens and other community members about our 2019 inaugural camp helped define the direction of BioJam in 2020. For example, Dr. Bryan Brown, a professor in the Stanford Graduate School of Education and an expert in science education in urban communities, visited the camp in 2019 and drew several connections with his own scholarship in youth-empowered science exploration. In his book *Science in the City*, Dr. Brown discusses a 'Generative Learning Principal' and cognitive apprenticeship that greatly informed and refined our approach. We moved towards a camp framework that started with a collectively-grown vocabulary list with teens as an entry point to invite them to grow their own scientific definitions from their lived experiences and existing knowledge. Then, using those anchor points, we expanded to include scientific terms and definitions, relying on iterative experimentation and language expansion to build deeper understanding. By removing the jargon and definitions created by an exclusionary science culture and instead leading with collective experiences as we worked on projects together, we sought to enable teens to move forward leaning into their own expertise as a starting point for engaging with science. Campers engaged in cognitive apprenticeship by experimenting with learning modules and deciding what to bring forward. Finally, the campers assembled their vocabulary and learning materials into workshops for their communities.

After the summer ended, the 2019 cohort continued to communicate and collaborate in monthly meetings to design, plan, and organize community workshops. The first series of workshops stemmed from a camp research presentation by Stanford student Alexa Garcia (Department of Bioengineering, Stanford University) about the ability of mealworms, specifically their stomach enzymes, to digest polystyrene plastics. This work inspired the teens to share with their communities through their own voices. Through hours of brainstorming and discussion, the teens developed a workshop that invited participants to explore the applications of mealworms in addressing climate change and plastic pollution. They first conducted the workshop, "Mealworm Chefs," at the Tech Interactive, a science museum in San Jose, CA (Fig. 2b). Workshop participants, ranging from kids to adults, were prompted to create a mealworm meal out of polystyrene container scraps and place them in the collaborative table displays of live mealworms digesting those creations. During a similar workshop at the Lawrence Hall of Science, a science museum in Berkeley, CA, the cohort invited youth to craft their own polystyrene creatures to take home along with instructions on acquiring and caring for mealworms.

Another community engagement project, "MycoQuilts," grew from the biomaterial activities in the summer camp and was heavily supported by Corinne Okada Takara. Teens drew from their own creativity and culture to design mold forms in TinkerCAD (Computer Aided Design), which they 3D printed and vacuum formed. In each mold, teens grew mycelium, the interconnected cell network of filamentous fungus, which they fed various substrates they had brought from home, such as ramen noodles and flowers. Together, each teen's mycelium 'bio-textile' was displayed in the Shriram Center for Bioengineering and Chemical Engineering at Stanford University (Fig. 2c). Additionally, the MycoQuilt and two of the teens traveled to the Construct 3D Printing and Digital Fabrication conference (Rice University, TX) to present the project (Fig. 2d). The collaborative textile was displayed with a poster the BioJam teens co-designed and co-presented. The teen project won a First Place poster prize amongst a field of undergraduate and graduate students. At the end of 2019, the BioJam program and teens had grown together through the many projects which fostered a sense of community and empowerment.

*Our 2020 journey*

Going into the summer of 2020, the world was hit by the COVID-19 pandemic. Camp was redesigned from being in-person to remote on Zoom. Each camper received an elaborate kit for bio-making (SI Fig. 2a), which was delivered to each teen's' home. In direct response to the pandemic, the camp focused on culturally-centered design for personal protective equipment (PPE). Teens designed speculative PPE for their favorite movie characters, other planets, and their own communities. Over a span of two weeks, the teens prototyped their PPE designs, engaged in group discussions on Zoom, and embedded biosensors, circuitry and coding into PPE. BioJam teens learned with guest speakers, who presented about their experience with agricultural, sports, and fashion-based PPE. The camp curriculum emphasized sustainability and biomaterials in prototyping, using paper, cardboard, pipe cleaners, and natural materials like kombucha leather and mycelium (Fig. 2e). Given the backdrop of the pandemic, we emphasized personal wellness and self-care. Teens participated in a quantitative evaluation at the end of camp using Google Forms developed in consultation with the Center for Teaching and Learning (CTL) at Stanford University.

Following summer camp, the 2020 BioJam team continued meeting with teens over Zoom to envision safe community engagement activities, which were sorely needed given the pandemic. The first community engagement activity was an art installation in the *Holding the Moment* exhibition at the Norman Y. Mineta San Jose International Airport (Fig. 2f). In response to the theme, "the impact of COVID-19 in your community," BioJam members reflected on their self-care practices, cultural health practices, and family life in the context of the pandemic by decorating face masks. Teens created art on templates used to grow mycelium face masks during camp (Fig. 2e) The BioJam piece was selected from over 300 entries and was the only teen submission out of the 77 artists featured. Following this exhibit, in collaboration with Xinampa (Salinas, CA) and the San Francisco Opera Costume Department, the BioJam teens organized a drive-up sewn and KN95 mask distribution for agricultural and labor workers in the Gonzales, CA area (Fig. 2g). Overcoming significant logistical obstacles, teens organized the entire event, from obtaining city permitting for the site, designing informational cards in four languages, creating original artwork, and meeting in-person to distribute the masks. Finally, teens' reflections on the pandemic, through original artwork, poetry, and prose, was published in Ginkgo Bioworks' *Grow*

Magazine, as part of their Pandemic Diaries series (*18*) (Fig. 2h). These diaries are included in the resource toolkit for Concentric by Ginkgo, a COVID-19 pool testing platform that serves over one thousand schools.

BioJam teens also presented at the 2021 Annual Meeting for the Society for Integrative and Comparative Biology (SICB) (Fig. 2i). Teens prepared and presented both a talk about BioJam Camp and an interactive workshop about biomaking (*19*, *20*). The workshop was based on materials developed by BioJam teens in the 2021 cohort ("Grow-Your-Own (GIY) BioBuddies"), and focused on imagining sustainable fashion using biomaterials (*21*). Teen instructors grew and dried kombucha leather (bacterial cellulose) at home for several months and contributed silhouettes which were laser cut for the activity. Kits were prepared and shipped to participants all over the world, which spanned children, undergraduates, graduate students, and academic faculty. Over Zoom, the teens deftly led the interactive talk and workshop, which used break-out sessions, interactive prototyping, and a virtual whiteboard to engage participants.

*Our 2021 journey*

Our 2021 camp focused on the themes of climate change, environmental justice, and sustainability. The two-week camp used a hybrid model, with the first day of camp held in-person at a community garden in Natividad Park (Salinas, CA) and an outdoor community space, Marina Park (San Leandro, CA). These sites were selected to facilitate in-person interaction with our youth who lived in different parts of the Greater Bay Area (Fig. 2j). At these in-person meet-ups, teens received kits with all materials needed for camp, including a greenhouse-like structure for imagining humans' relationship with biology, live plants, circuitry materials, and other maker items (Fig. 2k, SI Fig. 2b). After the first two days, the remainder of the camp was held virtually on Zoom with community educators from Salinas and South Monterey County, East Side San Jose, and Oakland leading a camp learning module, guest speakers, and CoLABorators facilitating other activities. Teens programmed Micro:bit+ computers to illuminate and water plants, planted Indigeneous seeds, contemplated thoughtful stewardship of stolen land, and considered sustainable agricultural practices by developing a hydroponics system (Fig. 2j). Teens created cyanotype art pieces of themselves with local plants. Each activity was scaffolded through a paper journal (Fig. 2l), which was supported by digital uploads and comments using virtual whiteboards. Camp ended with a showcase of the teens' work to community advisory board members, academics, and community members, who provided feedback on the teens' prototypes. The 2021 camp assessment was developed by an interdisciplinary team of researchers and students; student testimonials are featured in Fig. 3.

*Our local impact*

We view BioJam as far more than a summer camp to get teens engaged in science. As co-learners, co-designers, and CoLABorators, we think critically about our position, privilege, and power, and how we can leverage this power towards justice. We seek to expand our work without compromising our values, and share with other communities beyond our immediate reach.

Locally, we hope BioJam grows even more deeply embedded into interdisciplinary, intergenerational, and cross-cultural community spaces. Over the past year, the creation of BioJam CoLABS has expanded our organizational capacity. The BioJam CoLABS structure takes inspiration from non-hierarchical emergent organizations, such as ant colonies (*22*), where participants step up and step back into roles based on interest, availability, and need. We have tried to position CoLABS as a unique space at Stanford that models co-learning and collaborating *in* community.

Now, we are focusing on future growth being centered in Salinas, South Monterey County, East Side San Jose, and Oakland, CA. First, BioJam's Community Advisory Board is composed of organizations rooted in these communities. These organizations guide BioJam's vision and direction alongside former teen participants and mentors, who make up the Teen Advisory Board. We hope to grow these advisory boards as cross-generational, cross-community spaces for visioning, not only the future of BioJam, but the future of biodesign and innovation in the Greater Bay Area. Second, we would like to deepen the involvement and expertise of a broader network of community educators. Centering lived experience and expertise from our teens' communities and cultures is fundamental to BioJam's mission. We envision expanding our team of community educators and deepening their involvement throughout the academic year in curriculum development and supporting teen engagement projects.

Further, we hope to inspire other communities beyond the Greater Bay Area of California to collaborate in our vision for just, inclusive, and liberatory biodesign and bioengineering. We share our work through academic outlets such as conferences and papers, as well as venues like social media and mainstream print media. To encourage conversation and build community around biodesign education, we make our curriculum open-source and adaptable using [Creative Commons licensing](https://biojamcamp.weebly.com/showcase.html) (see https://biojamcamp.weebly.com/showcase.html). In the spirit of design, the structure of BioJam itself serves as a prototype not only for us, but also the broader biomaking community: adapting, growing, evolving.

*Our vision for the future*

More than a camp or an organization, BioJam represents a vision for the future. We dream of a world where bioengineering and biodesign starts in kitchens, alongside roads, in agricultural fields, and empty lots. We imagine cultural and religious practices, justice and lived experience, guiding how we develop technology— what technology is and can be.

The vibrance of the BioJam ecosystem comes from its diversity: artists imagine alongside scientists, academics learn with community activists, youth and elders collaborate to imagine a more just world. The structural change we seek comes from the relationships we build. Our camp is the culmination of shared connections: we work alongside each other in community gardens, commune over shared meals, mourn losses and celebrate victories. In centering mutual aid and interdependence, we hope that BioJam will continue to grow from a place of passion, vision, and imagination.

The transformative legacy of BioJam comes from its transience. As our youth grow, we imagine they carry their cultural identity as central to their praxis and vision as scientists, artists, and activists. Our community partners, educators, and academics leave BioJam being challenged to listen carefully, confront the Other, and work together to actualize the dreams of our youth. We hope they (and we) take that experience with them to galleries, classrooms, protests, conferences, and beyond. We hope that BioJam can be a seed from which we sow the gardens of the future.

**Figures:**

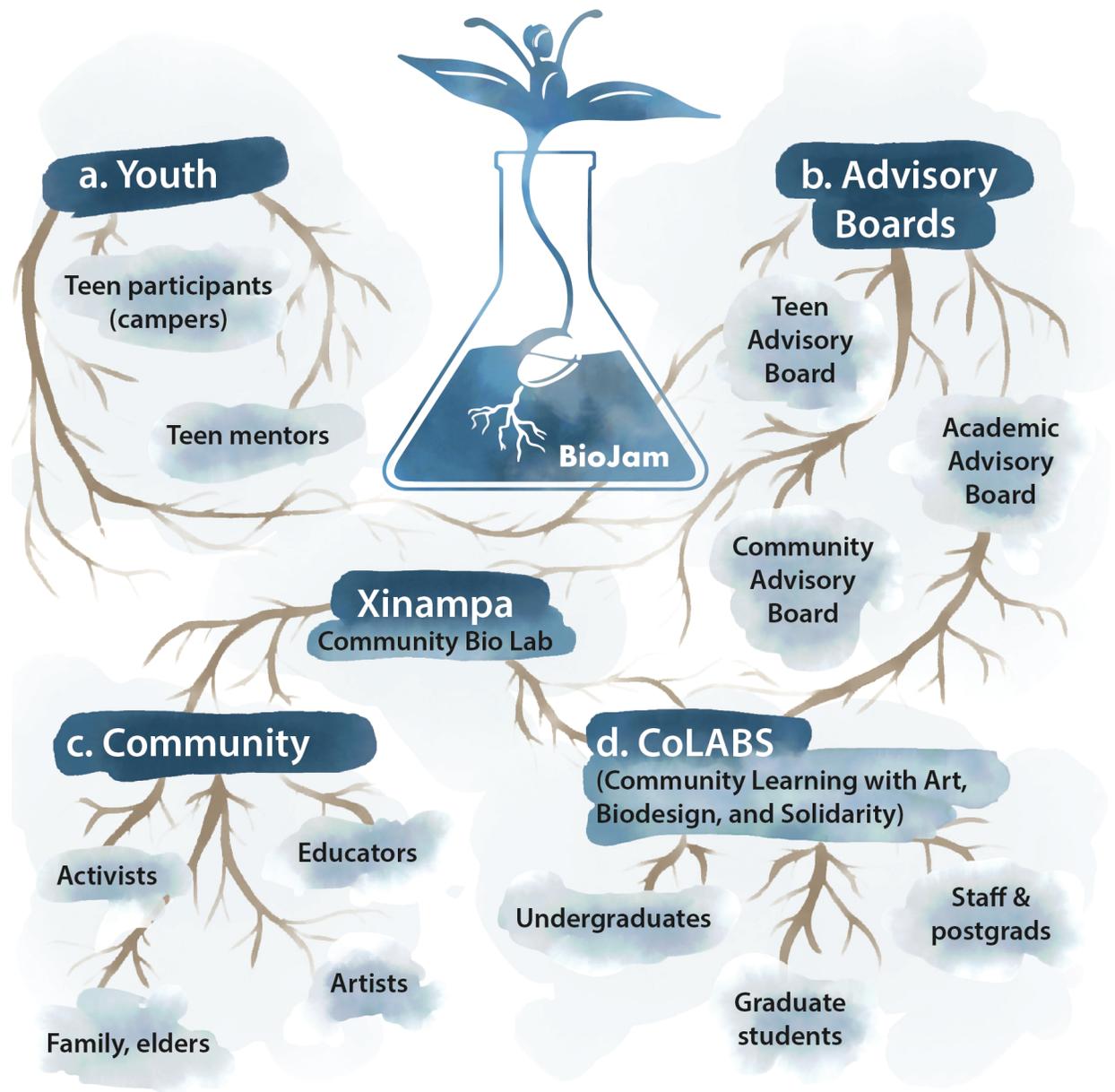

**Figure 1: The BioJam ecosystem**

The BioJam ecosystem is rooted in a web of supporting communities. (a) Starting with youth, BioJam's vision and direction is to engage teen participants (campers) and is guided by former teen participants who return as teen mentors. Teens also support the BioJam Advisory Boards (b), which also include community and academic members. These advisory boards guide the camp direction and programming, which is deeply supported by community educators (c). Primarily centered through Xinampa, a community bio lab based in Salinas, CA, our community educators develop our programming and curriculum. BioJam is coordinated by Stanford student organization BioJam CoLABS (Community Learning with Art, Biodesign, and Solidarity) (d), which is composed of undergraduate and graduate students, as well as post-graduates and staff.

**Alt text:** Image of a BioJam logo, which is a Erlenmeyer flask with a seed growing out of it into a flourishing human-like sprout. The text matches the figure legend, with each word in a blue bubble, connected with abstract roots/mycelium network.

## Summer camp

*Summer*

## Community engagement

*Academic year*

### 2019

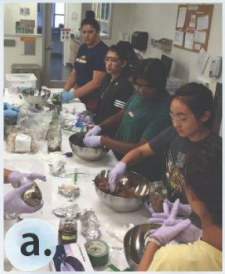
a.

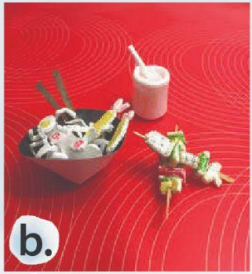
b.

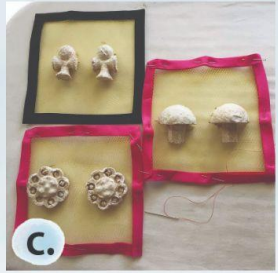
c.

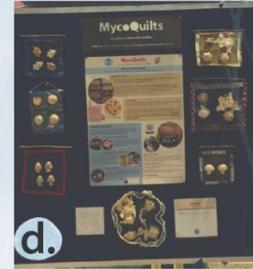
d.

### 2020

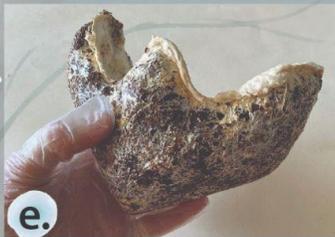
e.

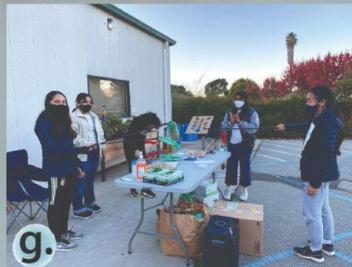
g.

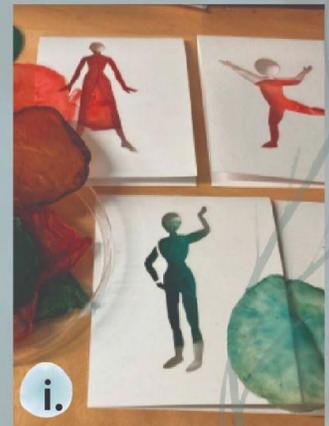
i.

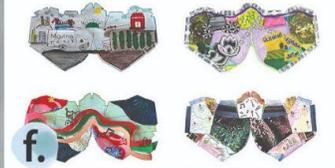
f.

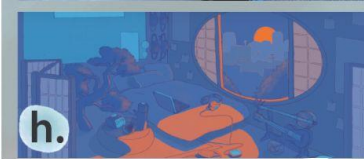
h.

### 2021

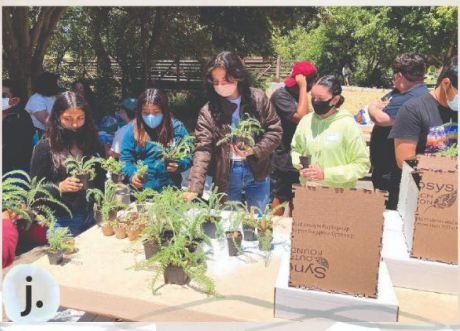
j.

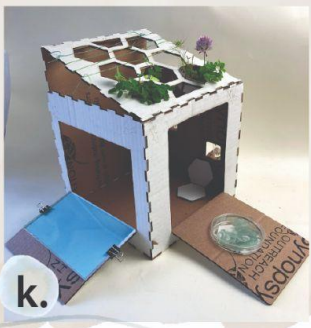
k.

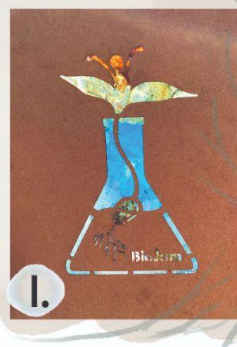
l.

**Figure 2: BioJam camp and community engagement projects**

In 2019, (a) teens grew mycelium in various substrates during summer camp. (b) Participants created styrofoam "meals" for plastic-digesting mealworms at a teen-designed BioJam workshop held at the Tech Interactive (San Jose, CA) and the Lawrence Hall of Science (Berkeley, CA). (c) Teen grew mycelium quilt pieces for the MycoQuilts project, presented at Stanford University and Construct 3D Printing and Digital Fabrication conference (Rice University, TX) (d). In 2020, teens explored various biomaterials for designing personal protective equipment (PPE) including mycelium (e). After camp, teens exhibited original artwork reflecting their experiences of the pandemic at the *Holding the Moment* exhibition at the Norman Y. Mineta San Jose International Airport (f), organized a mask distribution for agricultural and labor workers in Gonzales, CA (g), contributed original artwork, poetry, and prose to Ginkgo Bioworks' *Grow* Magazine (h, art originally published in *Grow* by Ginkgo), and led a workshop on biomaterials and fashion for the Society for Integrative and Comparative Biology (SICB) Annual Meeting (i). In 2021, teens explored climate change, sustainability, and environmental justice through in-person meet-ups (j), designing sustainable structures (k), and documenting their work in maker journals (l).

**Alt text:** At the top of the figure is a line going from the summer camp into community engagement projects. There are a series of images representing the descriptions in the figure legend, progressing from 2019 to 2020 to 2021, as described in the figure legend. Light green abstract shoots/mycelium weave the years together.

By being a part of the BioJam community since 2019, I was able to build upon my prior knowledge of biology and connect this interest to interdisciplinary fields such as engineering, design, and cultural history. Doing so allowed me to better serve my community in a meaningful way. It's amazing to see the program grow in a way that will truly benefit and encourage teens to get involved with biodesign.
—*Trisha* (2019 teen participant, 2020 teen mentor)

Being a BioJam enrollee helped me understand the importance of environmental justice and also how to incorporate bioengineering into our daily lives. It helped me discover my interest in Biology as well.
—*Sonam* (2021 teen participant)

BioJam showed me that I have a voice and I can do so much to help out not only my community but the world! Being a camper in Biojam in 2021 it has been so much fun and learning from other campers and teen mentors. I learned so many new things in the science field that I did not know and they encouraged me to keep on trying and not give up. —*Emily B* (2021 teen participant)

Biojam was a positive learning experience that increased my confidence in both prototyping and exploring new scientific concepts. I participated in the 2019 pilot program and served as a Teen Mentor in 2020, in both years appreciating how BioJam's focus on empowering teens to give voice to their communities makes this program truly meaningful and unique.
—*Anne* (2019 teen participant, 2020 teen mentor)

Taking part of BioJam Camp has allowed me to grow as a person by helping me see how I can effect others and my community in a small but helpful way.
—*Cristian* (2021 teen participant)

Being involved with BioJam has opened new doors along with given me new opportunities in ways I could better my community. It has taught me that there is a connection between our culture and science in situations we may not have associated. BioJam has shown me that my voice and opinions matter in my community and teens like me have the power to actually influence and better our community and environment. BioJam has made me excited and optimistic about the future, it's made me eager to help out and change my community and environment. BioJam has give me an opportunity to see the career path I want to pursue and how I could apply what I learn to the environment around me. —*Ary* (2021 teen participant)

**Figure 3: Our teens speak.**

Testimonials about the impact of BioJam over the years from our teen participants are displayed against a cyanotype background, featuring self-portraits of Ary and Elvia, two 2021 campers. The cyanotype self-portrait activity, which incorporates plants found in the teens' neighborhoods, was designed by Corinne Takara for Xinampa. The full text of these and additional testimonials is also included in the Supplemental Information.

**Alt text:** Blue background with white silhouettes of two teens from their original artwork. White clouds are behind the text of several teen testimonials (full text of all testimonials in the supplemental information). Light blue tendrils representing the mycelium motif continues around each testimonial.

**Supplemental Information:**

## S1: Extended teen testimonials

Biojam was a positive learning experience that increased my confidence in both prototyping and exploring new scientific concepts. I participated in the 2019 pilot program and served as a Teen Mentor in 2020, in both years appreciating how BioJam's focus on empowering teens to give voice to their communities makes this program truly meaningful and unique.
 –Anne, 2019 teen participant, 2020 mentor, 2021 teen advisory board member

Being encouraged to share and reflect on our own perspectives really helped us stay conscious/focused on how to maintain connection between the BioJam program and our communities. In the 2019 BioJam program, we were really inspired and excited by the "superpower" of mealworm gut bacteria to breakdown polystrene(sytrofoam). We were encouraged to explore the possibilities of it, and in the end we wanted to share the knowledge through activities to get people to reimagine the waste cycle in a fun, creative way.
 –Emily T., 2019 teen participant and 2020 mentor, 2021 teen advisory board member

Biojam is such an amazing program that allows you to learn many new things and it gives you the opportunity to share out what you know and you get to experience things that you wouldn't experience.
 –Elvia B., 2021 teen participant

Taking part of BioJam Camp has allowed me to grow as a person by helping me see how I can effect others and my community in a small but helpful way.
 –Cristian F., 2021 teen participant

BioJam allowed me to actively work on social skills as well as work on personal goals. I participated as a camper in 2020 and became a team mentor in 2021 for the program. This amazing group has allowed me to see/hear multiple perspectives,thoughts and ideas from teens that want to be involved it changing the world for the better.
 –Penny, 2020 teen participant and 2021 mentor

By being a part of the BioJam community since 2019, I was able to build upon my prior knowledge of biology and connect this interest to interdisciplinary fields such as engineering, design, and cultural history.  Doing so allowed me to better serve my community in a meaningful way. It's amazing to see the program grow in a way that will truly benefit and encourage teens to get involved with biodesign.
 –Trisha, 2019 teen participant and 2020 mentor, , 2021 teen advisory board member

BioJam showed me that I have a voice and I can do so much to help out not only my community but the world! Being a camper in Biojam in 2021 it has been so much fun and learning from other campers and teen mentors. I learned so many new things in the science field that I did not know and they encouraged me to keep on trying and not give up.
 –Emily B, 2021 teen participant

Being involved with BioJam has opened new doors along with given me new opportunities in ways I could better my community. It has taught me that there is a connection between our culture and science in situations we may not have associated. BioJam has shown me that my voice and opinions matter in my community and teens like me have the power to actually influence and better our community and environment. BioJam has made me excited and optimistic about the future, it's made me eager to help out and change my community and environment. BioJam has give me an opportunity to see the career path I want to pursue and how I could apply what I learn to the environment around me.
	–Ary, 2021 teen participant

Being BioJam enrollee helped me understand the importance of environmental justice and also how to incorporate bioengineering into our daily lives. It helped me discover my interest in Biology as well.
	–Soyang, 2021 teen participant

Having being part of Biojam allowed me to learn more about biology as well as my community. I learned through experiences that I believe that only Biojam could provide. It allowed me to see a new perspective in the science world that I was not aware of before. Every day was a new learning path that I enjoy very much. Not only did I learn a lot, I also got to meet and collaborate with many amazing people.
	–Jennifer M., 2021 teen participant

**S2: 2020 and 2021 BioJam camp kits**

## a. 2020 kit

## b. 2021 kit

**Figure 1: The BioJam ecosystem**

The BioJam ecosystem is rooted in a web of supporting communities. (a) Starting with youth, BioJam's vision and direction is to engage teen participants (campers) and is guided by former teen participants who return as teen mentors. Teens also support the BioJam Advisory Boards (b), which also include community and academic members. These advisory boards guide the camp direction and programming, which is deeply supported by community educators (c). Primarily centered through Xinampa, a community bio lab based in Salinas, CA, our community educators develop our programming and curriculum. BioJam is coordinated by Stanford student organization BioJam CoLABS (Community Learning with Art, Biodesign, and Solidarity) (d), which is composed of undergraduate and graduate students, as well as post-graduates and staff.

**Alt text:** Image of a BioJam logo, which is a Erlenmeyer flask with a seed growing out of it into a flourishing human-like sprout. The text matches the figure legend, with each word in a blue bubble, connected with abstract roots/mycelium network.

**Figure 2: BioJam camp and community engagement projects**

In 2019, (a) teens grew mycelium in various substrates during summer camp. (b) Participants created styrofoam "meals" for plastic-digesting mealworms at a teen-designed BioJam workshop held at the Tech Interactive (San Jose, CA) and the Lawrence Hall of Science (Berkeley, CA). (c) Teen grew mycelium quilt pieces for the MycoQuilts project, presented at Stanford University and Construct 3D Printing and Digital Fabrication conference (Rice University, TX) (d). In 2020, teens explored various biomaterials for designing personal protective equipment (PPE) including mycelium (e). After camp, teens exhibited original artwork reflecting their experiences of the pandemic at the *Holding the Moment* exhibition at the Norman Y. Mineta San Jose International Airport (f), organized a mask distribution for agricultural and labor workers in Gonzales, CA (g), contributed original artwork, poetry, and prose to Ginkgo Bioworks' *Grow* Magazine (h, art originally published in *Grow* by Ginkgo), and led a workshop on biomaterials and fashion for the Society for Integrative and Comparative Biology (SICB) Annual Meeting (i). In 2021, teens explored climate change, sustainability, and environmental justice through in-person meet-ups (j), designing sustainable structures (k), and documenting their work in maker journals (l).

**Alt text:** At the top of the figure is a line going from the summer camp into community engagement projects. There are a series of images representing the descriptions in the figure legend, progressing from 2019 to 2020 to 2021, as described in the figure legend. Light green abstract shoots/mycelium weave the years together.

**Figure 3: Our teens speak.**

Testimonials about the impact of BioJam over the years from our teen participants are displayed against a cyanotype background, featuring self-portraits of Ary and Elvia, two 2021 campers. The cyanotype self-portrait activity, which incorporates plants found in the teens' neighborhoods, was designed by Corinne Takara for Xinampa. The full text of these and additional testimonials is also included in the Supplemental Information.

**Alt text:** Blue background with white silhouettes of two teens from their original artwork. White clouds are behind the text of several teen testimonials (full text of all testimonials in the supplemental information). Light blue tendrils representing the mycelium motif continues around each testimonial.